\documentclass[prb, aip, twocolumn,superscriptaddress, amsmath, amssymb]{revtex4} 			 
\usepackage{graphicx}
\usepackage{bm} 
\usepackage{dcolumn} 
\usepackage[colorlinks=true, citecolor=blue, linkcolor=blue]{hyperref}%
\usepackage[normalem]{ulem}  

\begin{document}

\title{Dynamics and control of nonradiative excitons - free carriers mixture in GaAs/AlGaAs quantum wells} 

\author{A. S. Kurdyubov}
\affiliation{Spin Optics Laboratory, St. Petersburg State University, Ulyanovskaya 1, Peterhof, St. Petersburg, 198504, Russia}

\author{A. V. Trifonov}
\affiliation{Experimentelle Physik 2, Technische Universitat Dortmund, 44221 Dortmund, Germany}
\affiliation{Spin Optics Laboratory, St. Petersburg State University, Ulyanovskaya 1, Peterhof, St. Petersburg, 198504, Russia}

\author{B. F. Gribakin}
\author{P. S. Grigoryev }
\author{I. Ya. Gerlovin}
\author{A. V. Mikhailov}
\author{I. V. Ignatiev}
\affiliation{Spin Optics Laboratory, St. Petersburg State University, Ulyanovskaya 1, Peterhof, St. Petersburg, 198504, Russia}

\author{Yu. P. Efimov}
\author{S. A. Eliseev}
\author{V. A. Lovtcius}
\affiliation{Resource Center ``Nanophotonics'', St. Petersburg State University, Ulyanovskaya 1, Peterhof, St. Petersburg, 198504, Russia}

\author{M. A{\ss}mann}
\affiliation{Experimentelle Physik 2, Technische Universitat Dortmund, 44221 Dortmund, Germany}

\author{A. V. Kavokin}
\affiliation{Westlake University, School of Science, 18 Shilongshan Road, Hangzhou 310024, Zhejiang Province,
China}
\affiliation{Westlake Institute for Advanced Study, Institute of Natural Sciences, 18 Shilongshan Road,
Hangzhou 310024, Zhejiang Province, China}
\affiliation{Spin Optics Laboratory, St. Petersburg State University, Ulyanovskaya 1, Peterhof, St. Petersburg, 198504, Russia}

\date{\today}

\begin{abstract}
   Dynamics of nonradiative excitons with large in-plane wave vectors forming a so-called reservoir is experimentally studied in a high-quality semiconductor structure containing a 14-nm shallow GaAs/Al$_{0.03}$Ga$_{0.97}$As quantum well by means of the non-degenerate pump-probe spectroscopy. The exciton dynamics is visualized via the dynamic broadening of the heavy-hole and light-hole exciton resonances caused by the exciton-exciton scattering. Under the non-resonant excitation free carriers are optically generated. In this regime the exciton dynamics is strongly affected by the exciton-carrier scattering. In particular, if the carriers of one sign are prevailing, they efficiently deplete the reservoir of the nonradiative excitons inducing their scattering into the light cone. A simple model of the exciton dynamics is developed, which considers the energy relaxation of photocreated electrons and holes, their coupling into excitons, and exciton scattering into the light cone. The model well reproduces the exciton dynamics observed experimentally both at the resonant and nonresonant excitation. Moreover, it correctly describes the profiles of the photoluminescence pulses studied experimentally. The efficient exciton-electron interaction is further experimentally verified by the control of the exciton density in the reservoir when an additional excitation creates electrons depleting the reservoir.
\end{abstract}
\maketitle 

\section*{Introduction}

The optical excitation well above the energy of the lowest exciton transition in a quantum well (QW) leads to the generation of hot excitons and/or free electrons and holes. After a short time, if the sample temperature is low enough, $T << R_X/k$, where $R_X$ is the exciton Rydberg in the QW and $k$ is the Boltzmann constant, free carriers couple to form excitons. The hot excitons and those created by the carrier coupling can acquire large kinetic energies and efficiently propagate in the plane of the QW layer. The major part of these excitons that have large enough in-plane wave vectors $K_x$ forms a nonradiative reservoir. Note that if $K_x > K_c$, where $K_c$ is the wave vector of light in the QW plane, excitons cannot recombine with photon emission because the wave vector selection rule cannot be fulfilled. These excitons may live in a QW for tens of nanoseconds~\cite{Trifonov-PRB2015}.


The radiative excitons with small wave vectors that live within the light cone, $K_X < K_c$, are characterized by a very short radiative lifetimes of the order of 10~ps~\cite{Deveaud-PRL1991, Andreani-SSC1991, Akimov-PRB1997, Ivchenko-book2004, Khramtsov-JAP2016}. Therefore the areal density of the nonradiative excitons can exceed that of the radiative excitons by orders of magnitude in the steady-state experiments. It is important to note that the reservoir of nonradiative excitons can strongly affect the properties of the radiative excitons due to the exciton-exciton interactions. In particular, the exciton lines in optical spectra can be considerably broaden because of the scattering of the radiative and nonradiative excitons.
In the microcavity structures, the reservoir determines many important properties of the polariton radiation such as the threshold of the polariton laser, multistability, etc., see e.g., Refs.~\cite{Lagoudakis-PRL2003, Wouters-PRB2013, Takemura-PRB2015, Kavokin-book2017, Schmidt-PRL2019, Berger-PRB2020}.

The exciton coherence lifetime in high-quality quantum structures is also mainly limited by the scattering processes which can be caused by the reservoir of excitons, free carriers, and phonons. At low temperatures, however, the exciton-phonon interaction is considerably suppressed and the exciton-exciton and the exciton-carrier scattering play a major role in the exciton polarization dephasing. The presence of nonradiative (dark) excitons may be detrimental for certain applications in quantum technologies \cite{QT1, QT2}.

Photoexcited bright excitons radiatively recombine within first few tens of picoseconds. However, the exciton photoluminescence (PL) is typically observed during the much longer time, of the order of 1~ns~\cite{Damen-PRB1990}. This means that the PL is mainly sustained by the nonradiative reservoir, whose excitons are scattered into the light cone via collisions with other excitons and  quasiparticles such as free carriers and phonons. In fact, the dynamics of excitons in the reservoir governs the radiative properties of excitons in QW structures.

In a general case, the reservoir consists of a mixture of nonradiative excitons and free carriers~\cite{Gibbs-nphot2011, Bieker-PRL2015}. The dynamics of this system is determined by many fundamental processes such as the exciton and carrier energy relaxation and thermalization, the coupling of electrons and holes into excitons, dissociation of excitons at elevated temperatures, the scattering of the excitons from the reservoir into the light cone followed by their fast recombination. The rate of these processes strongly depends on the experimental conditions, namely, on the photon energy of the excitation beam, the excitation power density, the sample temperature. Many attempts have been made to study these processes experimentally and theoretically.

The experimental studies undertaken in the past were based on two main methods. The first one is the study of the PL kinetics~\cite{Damen-PRB1990, Eccleston-PRB1991, Yoon-PRB1996, Gurioli-PRB1998, Szczytko-PRL2004, Chatterjee-PRL2004, Deveaud-PRB2005, Deveaud-ChemPhys2005, Nakayama-JAP2015, Beck-PRB2016}. This is an indirect method of study of excitons in the reservoir because the PL appears only after the last process, namely, the exciton scattering into the light cone.

Another, more direct, method of study of the exciton dynamics in the reservoir is based on the terahertz (THz) experiments~\cite{Cerne-PRL1996, Kaindl-Nature2003, Chatterjee-PRL2004, Galbraith-PRB2005, Kaindl-PRB2009, Ulbricht-RevModPhys2011, Zybell-APL2014}. The THz radiation can excite the transition between 1s and 2p exciton states. Therefore, the resonant absorption at the corresponding frequency (typically, in the range from 1 to 2 THz depending on the QW width) can be used for measurements of the exciton concentration. The nonresonant THz absorption can be also used to measure the carrier density in the electron-hole plasma. The two-color pump-probe method, the near-infrared pump pulses, and the THz probe pulses are used to study the exciton and carrier dynamics in the reservoir. The drawback of this technique is its relatively low sensitivity related to the small cross-section of the THz absorption. Therefore, relatively large excitation powers of the pump pulses had to be used in these experiments.

Recently a new approach of tracking dark excitons in the reservoir with exciton polaritons was suggested~\cite{Schmidt-PRL2019}. This method is based on polariton bistability revealed the long-lived ($>$ 20 ns) exciton reservoir in microcavity exciton-polariton systems. 

The theoretical modeling of the dynamics of the excitonic reservoir is challenging as it may only rely on a scarce experimental data available. First attempts of this kind of modeling~\cite{Piermarocchi-PRB1996, Piermarocchi-PRB1997, Kira-PRL1998} are followed by an extensive and contradictory discussion of the major mechanisms of the formation of the exciton PL signal~\cite{Szczytko-PRL2004, Chatterjee-PRL2004, Deveaud-PRB2005, Deveaud-ChemPhys2005, Kira-PQE2006, Koch-nmat2006, Kaindl-PRB2009, Zybell-APL2014, Beck-PRB2016}. The widely used model assumes the formation of excitons in the reservoir by photoexcited electrons and holes. After being formed, these excitons are assumed to be scattered by other excitons or acoustic phonons into the light cone where they recombine by emitting light. Both processes, namely the exciton formation, and their scattering to the light cone, are slow compared to the exciton radiative recombination, because they involve interacions with acoustic phonons. The scattering to the light cone is further slowed down by a small number of quantum states available for excitons in the light cone as compared to the number of quantum states in the reservoir. These are the basic assumptions that explain the slow rise and decay of the exciton PL at the low excitation intensity and low temperature. However, the complex time dependence of the PL signal at the increased excitation power and/or elevated temperatures, requires a more sophisticated theory to be involved. New theoretical considerations are needed, in particular, to explain the shortening of the rise and decay times of the PL signal.

In Ref.~\cite{Kira-PRL1998} an explanation of the shortening of the PL rise time is proposed. The authors suggested that the PL signal at the exciton resonance frequency can be caused by the recombination of Coulomb-correlated electron-hole pairs rather than the excitonic recombination. This idea was experimentally tested in many works~\cite{Szczytko-PRL2004, Chatterjee-PRL2004, Deveaud-PRB2005, Deveaud-ChemPhys2005, Kaindl-PRB2009, Zybell-APL2014, Beck-PRB2016} however no general conclusion was drawn so far. The reasons for the persisting umbiguity regarding the origin of the exciton PL signal are: (1) there are too many processes are involved in the PL signal dynamics, (2) the rates of these processes are sensitive
to particular experimental conditions, (3) the experimental information is not sufficient to draw certain conclusions about each of the processes involved.

In the present work, we study the dynamics of nonradiative excitons and free carriers using the pump-probe spectroscopy of the nonradiative broadening of exciton resonances~\cite{Trifonov-PRL2019, Trifonov-Semicond2019}. This technique allows instantaneous access to the exciton and carrier concentrations in the reservoir, thus enabling the study of their dynamics. To distinguish between the contributions of the reservoir excitons and free carriers to the broadening dynamics of the exciton resonances, we study these dynamics at different energies of optical excitation. We found that, if the optical excitation creates both excitons and free carriers in the QW, the excitons in the nonradiative reservoir survive up to the next laser pulse. This means that their lifetime exceeds 12 ns. The small depth of potential wells for electrons and holes in the QW structure
under study allowed us to realize the excitation conditions where either electrons or holes or both types of carriers are created in the barrier layers. We found that the exciton reservoir is effciently depleted if there is an imbalance of the photocreated electrons and holes in the QW. This is a clear indication of the important, probably the key role of the exciton-carrier scattering in the depletion of the reservoir.

The paper is organized as follows. In the first two sections, we explain the employed method of detection of the excitation spectra of the nonradiative broadening of exciton resonances. This technique as well as the study of the photoluminescence excitation (PLE) spectra allow us to formulate a qualitative model of the energy structure of the exciton and carrier states in the structure under study. It is described in the section III. This model allows us to carefully choose experimental conditions for the study of the exciton and carrier dynamics in the reservoir. The experimental study of the dynamics is described in the section IV. The next section describes the model of the dynamics, which we develop to explain the main features of the experimentally observed dynamics. The model, in particular, predicts the PL signal profiles at different excitation conditions. We compare these predictions with experimentally measured PL kinetics in the section VI. In the section VII we discuss the obtained experimental results and their modeling. Then we demonstrate a possibility to control exciton reservoir in the section VIII. Finally, we summarize our main results in the last section.

\section{Reflectivity spectra}
\label{experimental}
The structure under study was grown by the molecular beam epitaxy (MBE) on n-doped GaAs substrate with a crystallographic orientation (001). It contains a 14-nm GaAs QW layer sandwiched between Al$_x$Ga$_{1-x}$As barrier layers characterised by a low content of aluminum, $x \approx 3$~\%. The low aluminum content allowed to maintain a high structural quality of samples, possibly due to the suppression of build-in deformations~\cite{Shapochkin-PRA2019}. The potential wells for carriers in this structure are shallow, a few tens of meV. In our experiments, we took advantage of the shallow potentials in order to study the energy structure of excitonic and carrier states in the whole energy range from the lowest heavy-hole exciton state in the QW up to the bulk exciton state in the barrier layers.

The reflectivity spectra of the structure were measured at the normal incidence of a probe laser beam. Spectrally-broad 80-femtosecond pulses of a Ti:sapphire laser were used in these experiments. They allowed us to cover the whole spectral range studied. The reflected laser beam was dispersed in a 0.5-meter spectrometer with a 1800 gr./mm grating and detected using a nitrogen-cooled CCD array. The normalization of the detected signal on the laser beam spectral profile was used to obtain reflectivity spectra of the sample.
\begin{figure}
   \includegraphics{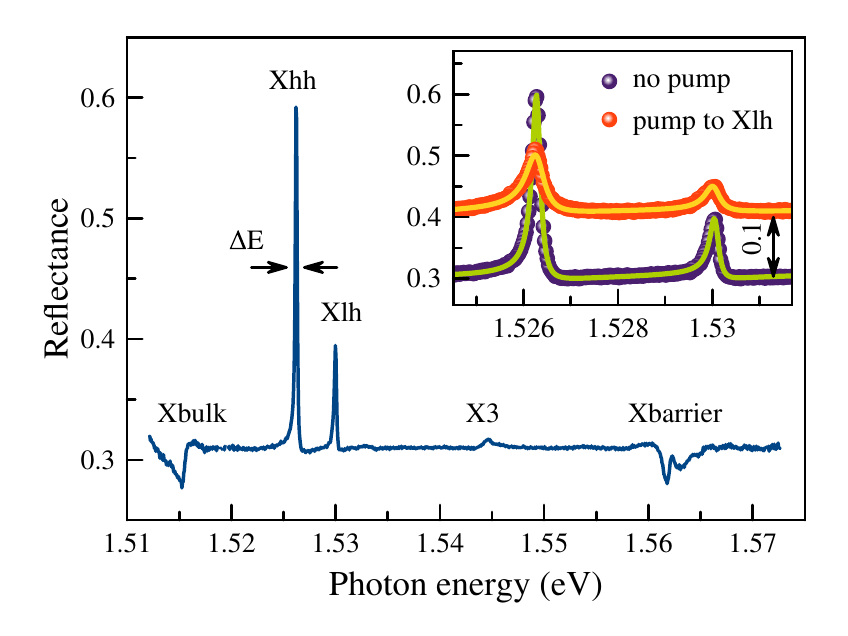}
   \caption{The reflectivity spectrum of the studied QW structure measured at $T = 4$~K. Labels Xhh and Xlh mark the ground states of the heavy-hole and light-hole excitons in the QW. Resonance ``X3'' can be tentatively ascribed to the third quantum-confined state of the heavy-hole exciton. Inset shows fragments of the reflectivity spectra measured with no pump (lower curve) and in the presence of an additional CW excitation of the power density $P_{\text{CW}} = 1$~W/cm$^2$ resonant with the Xhh exciton transition (upper curve). The lupper spectrum is vertically shifted by 0.1 for the clarity of presentation. Solid lines show the fits using Eqs.~(\ref{Eq1}, \ref{Eq2}).}
   \label{Fig1}
\end{figure}

A typical reflectivity spectrum of the studied QW structure is shown in Fig.~\ref{Fig1}. It displays the
the heavy-hole (Xhh) and light-hole (Xlh) exciton resonances in the spectral region of 1.525 -- 1.535~meV. The resonances appear as peaks due to the appropriate choice of the top layer thickness.
The reflectivity spectrum in Fig.~\ref{Fig1} also displays features related to the bulk excitons in the GaAs buffer layer and in the AlGaAs barrier layers. The energy distance between these features allows one to estimate the sum of the depth of potential wells for electrons and holes in the QW to be about 46~meV.

The inset in Fig.~\ref{Fig1} shows a fragment of the reflectivity spectra in the spectral vicinity of Xhh and Xlh resonances.
The spectral widths of the resonances are very sensitive to the excitation conditions. In the regime of a very weak probe beam intensity of the order of one $\mu$W per spot area of $10^{-4}$~cm$^2$ and no additional illumination applied, the full width at a half maximum of the Xhh resonance decreases down to $\Delta E = 185$~$\mu$eV. Such a small spectral width of an exciton resonance indicates a high quality of the sample. In the presence of an additional illumination by a CW laser, the resonances become broader, see the upper spectrum in the inset. The CW excitation creates excitons and/or free carriers so that the resonance broadening reflects the interaction of the bright excitons with the photocreated quasiparticles. This interaction can be monitored this way and we use it as the main method of study in our work.

The exciton resonances can be modeled in the framework of the nonlocal optical response theory described in Ref.~\cite{Ivchenko-book2004} and applied to the experimental data analysis in many works, see, e.g., Refs.~\cite{Trifonov-PRB2015, Trifonov-PRL2019, Grigoryev-SuperMicro2016, Shapochkin-PRA2019, Khramtsov-JAP2016, Khramtsov-PRB2019}. The amplitude reflection coefficient of a QW, $r_{\text{QW}}(\omega)$, in the vicinity of frequency $\omega_X$ of a single exciton resonance is given by the expression:
\begin{equation}
   \label{Eq1}
   r_{\text{QW}}(\omega) = \frac{i \Gamma_{\text{R}}}{\omega_X - \omega - i (\Gamma_{\text{R}} + \Gamma_{\text{NR}})},
\end{equation}
where $\Gamma_{\text{R}}$ and $\Gamma_{\text{NR}}$ are the radiative and nonradiative exciton relaxation rates, respectively. The total intensity of the reflected light depends also on the amplitude reflection coefficient of the sample surface, $r_s$, and can be expressed as:
\begin{equation}
   \label{Eq2}
   R(\omega) = \left|\frac{r_{s}+r_{\text{QW}}(\omega)e^{i2\phi}}{1+r_{s}r_{\text{QW}}(\omega)e^{i2\phi}}\right|^2,
\end{equation}
where $\phi$ is the phase acquired by the light wave propagating through the top layer of the
structure to the middle of the QW layer.

Eqs.~(\ref{Eq1}, \ref{Eq2}) are used to fit the spectra shown in the inset of Fig.~\ref{Fig1}. One can see that the calculated curves perfectly reproduce all peculiarities of the resonances. In particular, they describe the Lorentz-like profile of the resonances with slowly decaying wings. This is a clear evidence that no noticeable inhomogeneous broadening is present in the exciton system. Such broadening would result in the Gaussian-like wings characterized by a faster decay than the observed Lorentzian wings.

The good agreement of the experimental and calculated curves allows one to extract the main parameters of exciton resonances, namely, the radiative ($\hbar\Gamma_{\text{R}}$) and nonradiative ($\hbar\Gamma_{\text{NR}}$) broadenings as well as the exciton energy ($\hbar\omega_X$) with an accuracy from few $\mu$eV to fractions of $\mu$eV. Of course, some systematic errors in the obtained values are also possible. 

The homogeneous nonradiative broadening of exciton resonances, $\hbar\Gamma_{\text{NR}}$, is caused by the interaction (scattering processes) of bright excitons with other quasi-particles in the system, such as phonons, radiative and nonradiative excitons, free carriers~\cite{Trifonov-PRB2015, Shapochkin-PRA2019}. The effect of quasi-particles created by a CW illumination of the sample is demonstrated in the inset in Fig.~\ref{Fig1}. Interactions with these quasi-particles lead to increase of the nonradiative broadening of the Xhh resonance from 42~$\mu$eV up to 194~$\mu$eV at the relatively small CW excitation power of about 0.75~mW per laser spot $S_{\text{CW}} \approx 10^4$~$\mu$m$^2$. The high sensitivity of the nonradiative broadening to the scattering processes allows one to study in detail the energy spectrum of the quasi-particles involved and the dynamics of the scattering processes.

\section{NBE spectra}
\label{NBE}
To study the mechanism of nonradiative broadening, we additionally excited the sample by a tunable continuous-wave (CW) laser focused on the same point as the femtosecond probe beam. We tuned the energy of the CW laser, $\hbar\omega_{\text{CW}}$, in order to study the photoexcitation spectral dependencies of various characteristics of the Xhh and Xlh exciton resonances. The excitation power was stabilized at some fixed value, $P_{\text{CW}}$, by a variable attenuator with a build-in ``noise eater'' feedback feature. 
The reflectivity spectra were registered at wide range of $\hbar\omega_{\text{CW}}$ by the method described above and the obtained spectra were fitted using Eqs.~(\ref{Eq1}, \ref{Eq2}). As a result, the dependencies of all four parameters of each exciton resonance are obtained as functions of $\hbar\omega_{\text{CW}}$.

\begin{figure}
   \includegraphics{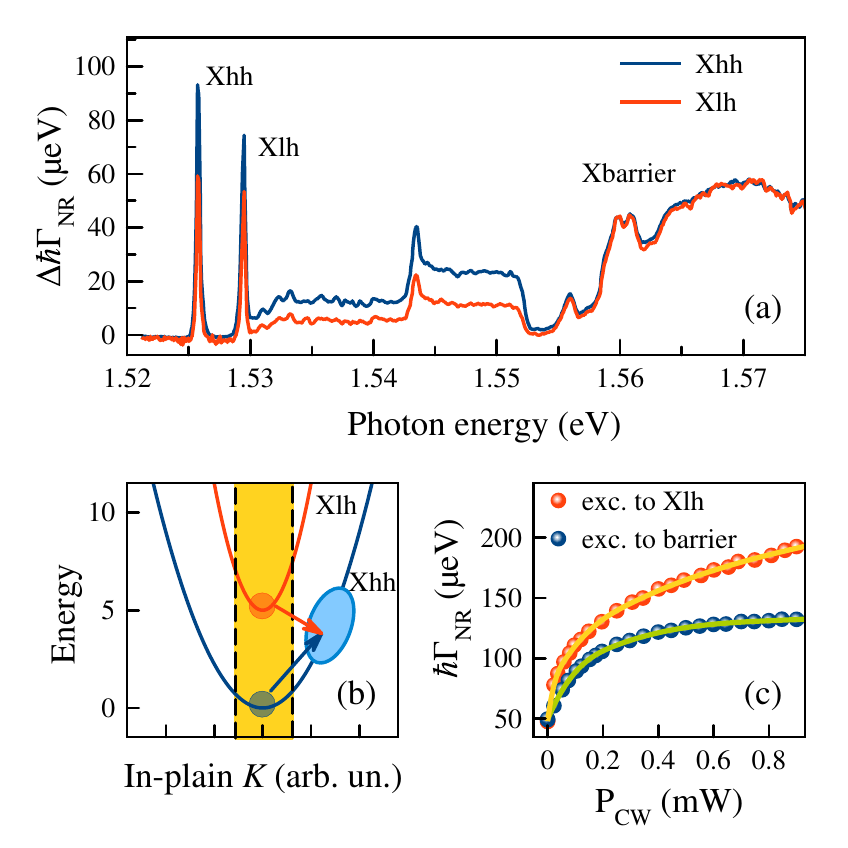}
   \caption{(a) Comparison of variable part of the NBE spectra for the Xhh (blue curve) and Xlh (red curve) exciton resonances. $P_{\text{CW}} = 0.15$~mW. The sample temperature $T = 4~K$. (b) Schematic representation of the population of the nonradiative reservoir. The parabolas show the Xhh and Xlh exciton energies as functions of the in-plane wave vector $K_x$ or $K_y$. The yellow area is the light cone and the light blue oval is the nonradiative reservoir. Arrows show the ejection processes.  (c) CW pump power dependencies of the nonradiative broadening of the Xhh resonance under CW excitation to the Xlh resonance ($E_{\text{CW}} = 1.530$~eV) and to the barriers ($E_{\text{CW}} = 1.570$~eV). Solid lines are the guides for the eye. 
   }
   \label{Fig2}
\end{figure}

Fig.~\ref{Fig2}(a) shows an example of the obtained dependencies of the nonradiative broadening $\hbar\Gamma_{\text{NR}}(\hbar\omega_{\text{CW}})$, for Xhh and Xlh resonances. Hereafter these dependencies will be referred to as the nonradiative broadening excitation (NBE) spectra. One can see that they demonstrate rich spectral structures, more pronounced than those of the reflectivity spectrum shown in Fig.~\ref{Fig1}. In particular, in the NBE spectrum, a strong increase of $\hbar\Gamma_{\text{NR}}$ is observed at the Xhh and Xlh resonances, which are also seen in the reflectance spectrum. Besides, a step-like increase of the broadening, an additional peak, and a dip are observed in the range of 1.53 -- 1.56~eV. These features can hardly be observed in the reflectivity spectrum.

The NBE spectrum of the Xlh resonance is similar to one of the Xhh resonance. The amplitude of the spectral features in the spectral range 1.525 -- 1.552~eV, however, is smaller than that in the NBE spectrum of the Xhh resonance. Fig.~\ref{Fig2} compares the variable part of the broadening, $\Delta \hbar\Gamma_{\text{NR}} = \hbar\Gamma_{\text{NR}} - \hbar\Gamma_{\text{NR}}^{(0)}$, where $\hbar\Gamma_{\text{NR}}^{(0)}$ is the broadening measured in the absence of the CW pumping: $\hbar\Gamma_{\text{NR}}^{(0)} = 42$~$\mu$eV for the Xhh resonance and $\hbar\Gamma_{\text{NR}}^{(0)} = 79$~$\mu$eV for the Xlh resonance. The larger value of $\hbar\Gamma_{\text{NR}}^{(0)}$ for the Xlh resonance can be associated to the exciton energy relaxation between Xlh and Xhh states.

The analysis of the data presented in Fig.~\ref{Fig2} allows us to draw several important conclusions. The features observed in the NBE spectrum are caused by the absorption of the CW radiation and generation of the excitons and/or free carriers. Excitons generated in the resonant absorption regime at the Xhh and Xlh transitions can recombine and emit photons. There exists, however, a competing relaxation process, namely, they can be ejected from the light cone and populate a reservoir of nonradiative excitons with large in-plane wave vectors~\cite{Trifonov-PRB2015}. This process is illustrated in Fig.~\ref{Fig2}(b). The ejection of the Xlh excitons can occur with the emission of acoustic phonons and, therefore, it is possible at zero temperature.

The ejection of the Xhh excitons can occur with absorption of phonons, which is possible at elevated lattice temperature. However, the critical kinetic energy of the Xhh excitons for the ejection is small, $E_c = (\hbar^2 K_c^2)/(2 M_{xy}) \approx 0.17$~meV.  Here $M_{xy} = m_0 (m_e^* + m_{hhxy}^*) = 0.177\, m_0$ is the in-plane exciton mass~\cite{Vurgaftman-JAP2001}, $K_c$ is the wave vector corresponding to the edge of the light cone, $K_c = 2\pi/(\lambda/n) \approx 2.8 \times 10^5$~cm$^{-1}$, where $n = 3.6$ is the refractive index of GaAs. The critical energy $E_c$ is achieved by the excitons at the sample temperature $T \approx 2$~K that is small enough. So, the Xhh ejection is possible at the temperatures used in our experiments.


The ejection efficiency depends on the ratio of the ejection and radiative recombination rates. It is larger for Xlh excitons as compared to Xhh excitons because of the smaller radiative decay rate and higher rates of processes associated with emission of phonons for the case of Xlh excitons. Consequently, the population of the exciton reservoir via excitation of the Xlh exciton resonance is more efficient than via excitation of Xhh excitons. This is why the Xhh and Xlh peaks in  Fig.~\ref{Fig2}(a) are almost of the same amplitudes, in contrast to those in the absorption spectrum (compare to Fig.~\ref{Fig1}).

Due to the large lifetime of the nonradiative excitons, the reservoir can accumulate a large amount of excitons. These are the quasi-particles, which are mainly responsible for the broadening of exciton resonances in the reflectivity spectra at these excitation conditions. The reservoir excitons have enough time to  thermalize at the temperature $T$ of the crystal lattice~\cite{Basu-PRB1992}. At low $T < 10$~K, used in the experiments, the kinetic energy of the excitons, $E_{\text{kin}} = kT < 1$~meV, that is smaller than the energy separation between the Xlh and Xhh exciton energies, $\delta E \approx 3.8$~meV, see inset in Fig.~\ref{Fig1}. Therefore the majority of excitons in the reservoir are the heavy-hole excitons.

A theoretical analysis shows~\cite{Ciuti-PRB1998} that the main mechanism of interaction of the reservoir excitons with the radiative excitons is the exchange interaction. The strength of interaction of Xhh reservoir excitons with Xhh radiative excitons (the Xhh-Xhh interaction) is composed by almost equal contributions of the electron-electron exchange interaction, $S_{e-e}$, and the hole-hole exchange interaction, $S_{hh-hh}$. The similar exchange interaction with Xlh radiative excitons (the Xhh-Xlh interaction) is approximately twice weaker because of the heavy-hole-light-hole exchange, $S_{hh-lh}$, is suppressed. Therefore we should expect that the photoinduced broadening of the Xlh resonance should be twice smaller than that of the Xhh resonance.
Experimental data presented in Fig.~\ref{Fig2}(a) qualitatively confirm this conclusion. In particular, the experimental ratio of the Xhh and Xlh NBE amplitudes [blue and red curves in Fig.~\ref{Fig2}(a)], $R \approx 1.9$, in the spectral region $\Delta E = 1.530 - 1.552$~eV. For the Xhh and Xlh peaks, this ratio is slightly less, $R \approx 1.6$.

Interestingly, the ratio $R$ is close to 2 in a wide spectral region up to 1.552~eV, where not only excitons but also free electrons and holes can be generated. In the higher energy region, the NBE spectra for the Xhh and Xlh resonances almost coincide. The observed difference in the behavior of NBE spectra in two spectral regions is a clear indication that the dynamics of quasi-particles is different in these regions. These states are discussed in the next section.

Panel (c) in Fig.~\ref{Fig2} shows the CW pump power dependence of the Xhh resonance broadening at the excitation into the Xlh and barrier optical transitions. The observed sublinear dependencies are the characteristic features of the complex dynamics of the exciton-free carrier reservoir discussed in Section~\ref{Kinetics-model}. 
Another possible reason for the sublinearity observed for the excitation to the Xlh resonance is related to the decrease of the absorption coefficient once the nonradiative broadening of this resonance increases~\cite{Ivchenko-book2004, Mursalimov}.

\section{PLE spectra}
\label{Model}

\begin{figure}
   \includegraphics{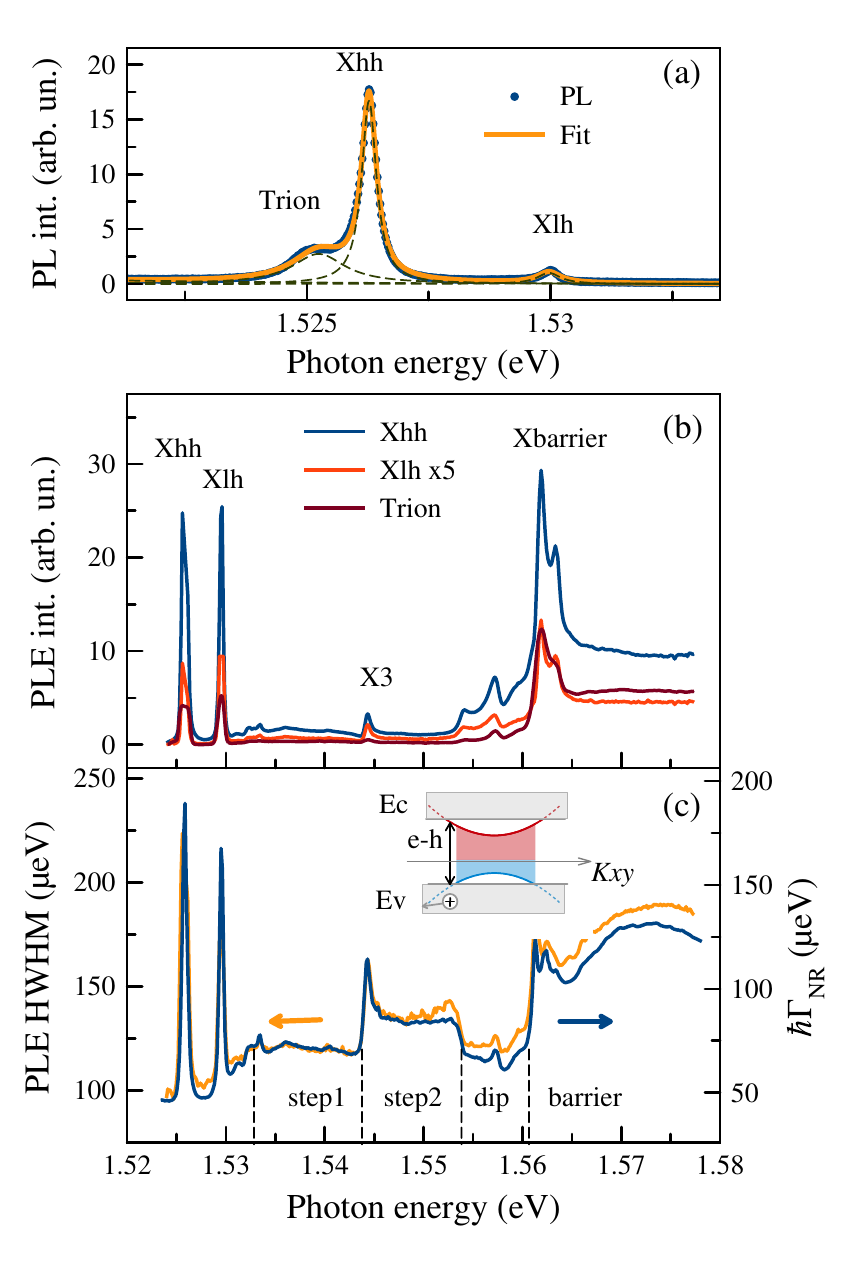}
   \caption{(a) An example of PL spectrum measured at the excitation to the barrier layer optical transitions ($E_{exc} = 1.58$~eV, $P_{exc}$ = 0.78~mW). Solid curve shows the fit by sum of Lorentzians (dashed curves). (b) PLE spectra of the Xhh, Xlh, and trion lines obtained as areas $S_j$ under respective Lorentz contours as functions of the photon energy of the CW excitation. The intensity of the Xlh line is multiplied by 5. (c) Comparison of the NBE spectrum (blue curve, right axis) and of the spectrum of HWHM (orange curve, left axis) for the Xhh exciton. The dashed vertical lines separate the spectral regions discussed in the text. The inset illustrate the mechanism of the ``dip'' spectral region formation.
   }
   \label{Fig3}%
\end{figure}

PLE spectra are typically used to study excited state of excitons and free carriers in heterostructures. We measured the PLE spectra using the same setup as for measurements of the reflectivity spectra (see Sect.~\ref{NBE}). However, the PL spectrum, rather than the reflectivity spectrum, was measured at each photon energy of the CW excitation.
Figure~\ref{Fig3}(b) shows a representative set of the PLE spectra of three spectral resonances observed in the PL spectrum. These are the Xhh and Xlh exciton lines as well as the third line, which is seen at the low energy side of the Xhh line. We shall comment on the nature of this latter resonance below. Each spectral line can be accurately modelled by a Lorentz function,
$$
   L(\omega) = \frac{S}{\pi\delta\omega[1+(\omega_0 - \omega)^2/(\delta\omega)^2]},
$$
where $S$ is the area under the function, $\omega_0$ is the resonant frequency, and $\delta\omega$ is the half width at the half maximum (HWHM) of the Lorentzian. Examples of the PL spectra and their fit by Lorentzians are shown in Fig.~\ref{Fig3}(a). The very good fit of the PL spectra by the sum of Lorentzians allows us to extract the main parameters of the spectral lines. The low energy spectral line can be tentatively ascribed to the trion transition shifted from the Xhh exciton by the coupling energy, $\delta E_T = 1.04 \pm 0.1$~meV. We believe that the observed feature can be a sum of spectral lines of the positively (X$^+$) and negatively (X$^-$) charged trions and it may also contain a contribution of bi-excitons.

The PLE spectra, measured as the dependence of the area $S_j$ on the photon energy of the CW excitation for three observed PL lines, are shown in Fig.~\ref{Fig3}(b). The spectra reveal several main peculiarities similar to those observed in the reflectivity spectrum (see Fig.~\ref{Fig1}). However in the ``dip'' spectral region, the spectra demonstrate {\it an increase} of the PL intensity for all three PL lines, rather than a decrease observed in the NBE spectra, see Fig.~\ref{Fig2}(a). This means that the absorption coefficient in this spectral region has no dip. This intriguing result becomes even more contradictory if we consider the excitation spectrum of the broadening of the PL lines. 

The Xhh PL line broadening, $\hbar\delta\omega_{\text{PL}}$, reveals a nonmonotonic dependence on the excitation energy. Fig.~\ref{Fig3}(c) compares this dependence for the Xhh exciton with the NBE spectrum. A remarkable similarity of both spectra is observed. In particular, a dip in the PL line broadening is clearly seen in the same spectral region as in the NBE spectrum. 

There is a systematic shift of the PL line broadening magnitude by a value of about 50~$\mu$eV relative to that of the NBE signal, as one can see comparing the left and right vertical axes in this figure. This shift can be explained by the contribution of radiative broadening of the exciton resonance, $\hbar\delta\omega_{\text{PL}} \approx \hbar\Gamma_{\text{NR}} + \hbar\Gamma_{\text{R}}$. A bit larger value of the shift than $ \hbar\Gamma_{\text{R}} = 37$~$\mu$eV (see the previous section) observed experimentally possibly points out to a systematic error in the measurements. 

The spectra shown in Fig.~\ref{Fig3}(c) can be divided into several spectral regions marked ``step1'', ``step2'', ``dip'', and ``barrier''. The region ``step1'' evidently corresponds to the optical transitions, hh1 -- e1, between the low-energy free electron (e1) and heavy-hole (hh1) quantum-confined states with finite wave vector along the QW layer. The density of states for the quasi-2D carriers with parabolic dispersion is known to be constant (see, e.g., textbook~\cite{Davies-book}) that explains the flatness of the spectra in this region. The left edge of the step is bound to the Xhh exciton and shifted from it by the exciton Rydberg energy, $R_x \approx 7$~meV~\cite{Khramtsov-JAP2016}. 

The interpretation of the ``step2'' plateau is more problematic because of the strong mixing of the hole states in QWs~\cite{Miller-PRB1985, Ivchenko-PRB1996, Zunger-PRB2000}.
The X3 peak can be tentatively ascribed to the third quantum-confined state of the Xhh exciton. The even symmetry of this state provides stronger exciton-light coupling than that of the second quantum-confined state Xhh2~\cite{Khramtsov-PRB2019}.

The spectral region marked as ``dip'' in Fig.~\ref{Fig3}(c) is the most intriguing part of the spectra. On one hand, the dip in the line broadening seems to indicate a smaller number of photocreated excitons and carriers, which scatter the bright excitons. This can be interpreted as a decrease of the absorption coefficient in this region. On the other hand, the PLE spectra in Fig.~\ref{Fig3}(b) demonstrate a strong increase of the PL intensity under excitation in this spectral region that points out to the increased absorption.

We assume that the ``dip'' corresponds to the optical transitions, which create carriers of a specific type, either electrons or holes, in the QW layer and the other type in the barrier layers. Under such circumstances, the nonradiative excitons cannot be accumulated efficiently in the reservoir as a large number of free carriers of one type scatter them effectively into the light cone. This should be clearly visible in the time-resolved measurements, so we do spectrally-resolved pump-probe experiments next. They are described in the following section.

A scheme of the optical transition under excitation to this region is shown in the inset in Fig.~\ref{Fig3}(c). The parabolic dispersion curves describe the carrier energies as functions of the in-plane carrier wave vector $k_x$. At some values of $k_x = k^{\text{max}}$, these curves cross the levels $V_c$ or $V_v$ (the solid horizontal lines) corresponding to the bottom of conduction band or the top of valence band in the barriers. These crossing points correspond to the delocalization of carriers in the barrier layers. 
This delocalization determine the left edge of the ``dip'' spectral region. The right edge is formed when both types of the carriers are delocalized in the barrier layers. 


The last spectral region marked in Fig.~\ref{Fig3}(c) as ``barrier'' is evidently formed by the optical transitions between the electron and hole states in the barriers. The density of states in the thick barriers is characterized by the square root energy dependence that explains a smooth increase of the signal. The signal decrease at the photon energies above 1.575~eV is possibly related to the light absorption in the heterostructure layers, which are far from the QW layer.

\section{Pump-probe experiments}
\label{dynamics-experiment}

To study the exciton dynamics, we have performed spectrally resolved pump-probe measurements. A femtosecond Ti:sapphire laser was used as a light source. The laser beam was split into the pump and probe beams. The pump beam was passed through an acousto-optic tunable filter forming spectrally narrow pump pulses of HWHM $\delta E_{\text{pump}} \approx 0.48$~meV. It was directed to the sample perpendicularly to its surface and focused into a spot with a diameter of about 95~$\mu$m. The delayed in time spectrally-wide probe pulses, $\delta E_{\text{probe}} \approx 45$~meV at the level 0.1 of their maximum, were used to detect the reflectivity spectra as it is described in section~\ref{experimental}. The measured spectra were analyzed with the use of Eqs.~(\ref{Eq1}, \ref{Eq2}).

\begin{figure}
   \includegraphics{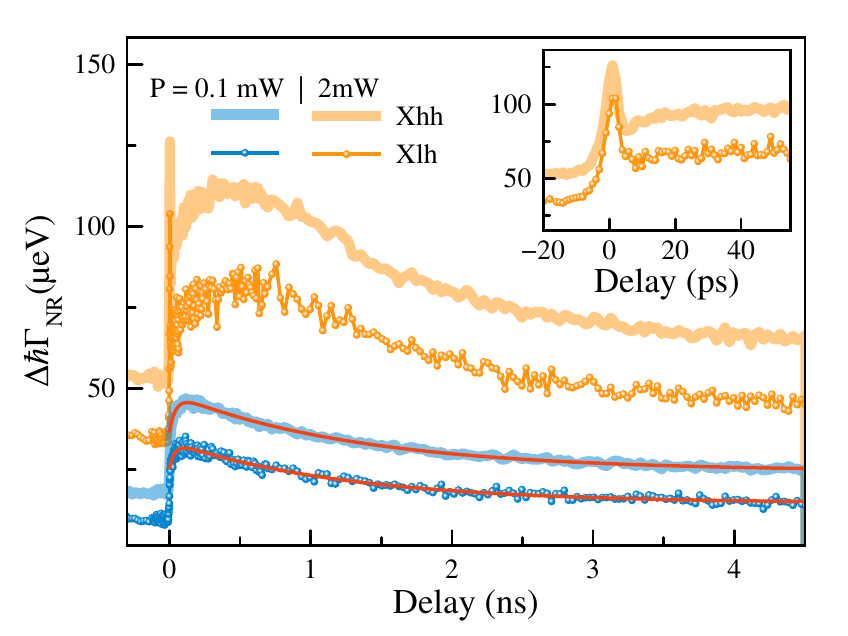}
   \caption{Delay dependencies of the nonradiative broadening of the Xhh (solid lines) and Xlh (circles) resonances obtained at the excitation to the Xlh transition with different excitation powers given in the legend. The background broadening not related to the excitation is subtracted. The smooth solid curves show the phenomenological fits of the Xhh and Xlh resonance broadenings by function~(\ref{Eq-fit}) with characteristic times $t_1 \approx 40$~ps and $t_2 \approx 1300$~ps. Inset shows initial parts of the experimental dependencies. The sample temperature $T = 4$~K.
   }
   \label{Dynamics1}
\end{figure}

Figure~\ref{Dynamics1} shows examples of dependencies of the nonradiative broadening of the Xhh and Xlh resonances on the time delay $\tau$ between the pump and probe pulses. The laser excitation energy was centered at the Xlh transition. The dynamic curves measured at the excitation to the Xhh transition demonstrate very similar behavior. In Fig.~\ref{Dynamics1}, only the variable parts of the broadening, $\Delta\hbar\Gamma_{\text{NR}}$, induced by the pump pulses are shown. The dependencies measured at the strong excitation power, $P = 2$~mW, consist of a very narrow initial part (pulse) with the duration of about several ps (see the inset to the figure) and a slowly varying signal. The dynamics of the signal may be ascribed to the interactions of radiative excitons prior to their radiative recombination that has a characteristic time of about 10~ps~\cite{Trifonov-PRB2015, Khramtsov-JAP2016}.

The slowly varying signal is characterized by the rising and decaying parts. At the weak excitation, the signal can be  approximated by a two-exponential function and a background,
\begin{equation}
\label{Eq-fit}
	f(\tau)=a \left(e^{-\tau/t_2} - e^{-\tau/t_1}\right)+b
\end{equation}
where $t_1$ and $t_2$ characterize, respectively, the rise and decay times of the dependence. The constant $b$ takes into account a very long-lived component of the signal with a characteristic lifetime exceeding the repetition period, $T_L = 12.5$~ns, of the laser pulses. This component manifests itself by a non-zero signal at the negative delay. Examples of the fit are shown in the figure by the smooth solid curves. There is an initial part of the dependence rising with a characteristic time $t_1$ of about 40~ps and a decaying part characterized by the time $t_2$ of about 1.3~ns. This behavior of the broadening reflects the complex dynamics of excitons in the reservoir.

If the excitation power is relatively low, the light-induced broadening of the Xlh resonance is noticeably smaller than that of the Xhh one, see blue curves in Fig.~\ref{Dynamics1}. It becomes approximately twice smaller at large delays. Such behavior is consistent with the broadening of the resonances observed in the steady-state experiments, see Fig.~\ref{Fig2}. The mechanism of the Xlh exciton broadening is already discussed in Sect.~\ref{NBE}.

The broadening of the Xhh and Xlh resonances sublinearly rises with the excitation power so that it is only of about 2.5 times larger at $P_{\text{pump}} = 2$~mW than that at the weak excitation, $P_{\text{pump}} = 0.1$~mW. This sublinear behavior is consistent with the results obtained at the CW excitation, see Sect.~\ref{NBE}. The ratio of the Xhh and Xlh broadenings decreases with the excitation power, in particular, at small delays. This observation possibly points out to some increase of the exciton reservoir temperature and, correspondingly, incomplete conversion of the photocreated light-hole excitons in the heavy-hole ones. The dynamics of the exciton resonance broadening is even more complex at the strong excitation because it cannot be approximated by the simple expression~(\ref{Eq-fit}). A model of the dynamics of formation of the excitonic emission signal is described below in Sect.~\ref{Xlh-model}.


In the case of excitation at higher energies up to the ``dip'' region, relatively small changes in the dynamics are observed. First, there is no rising part of the dynamics. Second, the broadenings of the Xhh and Xlh resonances becomes close in magnitude at the initial part even at the low excitation power. However, in the case of excitation to the ``dip'' and to the barrier exciton, the broadening dynamics of the Xhh and Xlh resonances almost coincides.


\begin{figure}
   \includegraphics{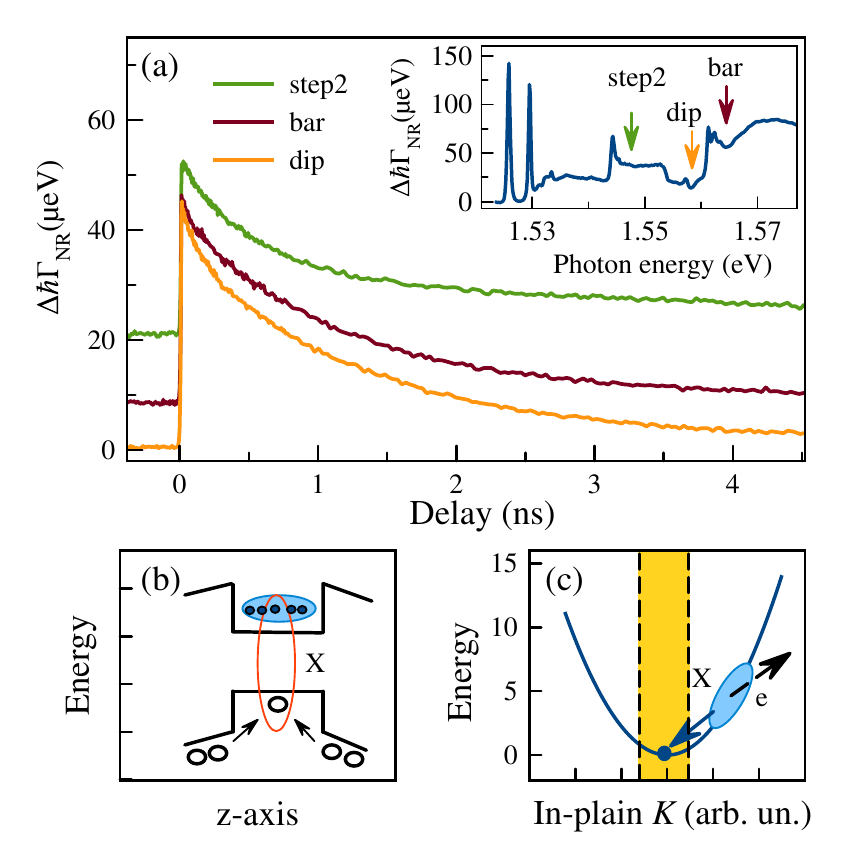}
   \caption{(a) Comparison of dynamics of the nonradiative broadening of the Xhh resonance obtained at the excitation to different spectral points indicated by arrows in the inset. The phonon-induced background is subtracted. Excitation power $P = 0.2$~mW. The sample temperature $T = 4$~K. (b) Illustration of the mechanism of formation of charged excitons under excitation into the ``dip'' spectral region. (c) The mechanism of fast depletion of the reservoir of the nonradiative excitons via the exciton scattering into the radiative Xhh state with a transfer of the excess momentum to a free carrier.
   }
   \label{Dynamics4}
\end{figure}


Fig.~\ref{Dynamics4}(a) shows the time-resolved dynamics of the Xhh resonance broadening under the excitation to several spectral points. At the first glance, the dynamics look similar. However, there is a drastic difference in the behavior of the long-lived component of the dynamics. The excitation to the ``step2'' or barrrier gives rise to the accumulation of the reservoir excitons so that the long-lived dynamic component survives up to the next laser pulse. This component is seen as the background signal at the negative delay. 

Similar behavior of dynamic curves is observed under excitation to the barrier layer optical transitions with higher photon energy. Such excitation created hot free electrons and holes in the barrier layers, which can run far away from the QW layer. Their coming to the QW layer is accompanied by several processes like, the carrier cooling and partial recombination in the barrier layers, the diffusion and the capture to the QW layer, etc. We therefore will no consider this case in detail.    

In contrast to those cases, the excitation to the ``dip'' is not accompanied by the accumulation of excitons in the reservoir. No noticeable signal is observed at the negative delay. As a result, the integral under the dynamic curve taken over the repetition period of laser pulses is considerably smaller than the similar integral under curves marked ``step2'' and ``bar''. This observation explains the dip in the NBE spectrum, see inset in Fig.~\ref{Dynamics4}(a).

The marked feature of the dynamics points out that there is an efficient mechanism of the depletion of the exciton reservoir if the structure is excited to the ``dip'' region. This mechanism cannot be a relaxation to the deep centers or other processes linked to the nonradiative losses of the optical excitations. Indeed, the PLE spectra shown in Fig.~\ref{Fig3}(a) demonstrate the {\it increase}, rather than the decrease, of the PL intensity under excitation to this spectral region. So, we have to conclude that the depletion mechanism is an efficient scattering of the reservoir excitons to the light cone followed by their radiative recombination.

We interpret this observations by a depletion mechanism illustrated in Fig.~\ref{Dynamics4}(b) and (c). According to the model of the energy structure discussed in the previous section, the optical excitation to the ``dip'' creates one type of carriers in the QW and the other one in the barriers. The accuracy of energy parameters for GaAs/AlGaAs heterostructures, such as the conduction and valence bands offsets as well the Luttinger Hamiltonian parameters for the valence band~\cite{Vurgaftman-JAP2001}, is insufficient to draw a definitive conclusion regarding the type of carriers that are first delocalized in the barrier layers. We assume that the holes are delocalized before electrons so that an excess electron concentration is present in the QW at the excitation to the ``dip'' spectral region, see Fig.~\ref{Dynamics4}(b). If a hole is captured from the barrier layers to the QW, it couples with one of the electrons, and an exciton is created in the nonradiative reservoir. However, this exciton is surrounded by many other electrons. Therefore the exciton can be scattered into the light cone while its excess momentum is transferred to one of the free electrons. This process is schematically shown in Fig.~\ref{Dynamics4}(c).

In the framework of this model, the reservoir mainly consists of electrons rather than excitons. The electrons equally interact with the photocreated heavy-hole and light-hole excitons that explains the equal broadening of the Xhh and Xlh resonances observed experimentally under excitation to the ``dip'', see Fig.~\ref{Fig2}(a). If electrons were delocalized in the barriers first, the reservoir would contain excess holes. The holes cool down to the lattice temperature and are converted to heavy holes. These holes can broaden the Xhh resonance via the exchange interaction but they cannot broaden the Xlh resonance. This contradicts the experiment.

\section{The model of the exciton dynamics}
\label{Kinetics-model}
\subsection{Excitation to the Xlh resonance}
\label{Xlh-model}

We assume that the resonant excitation to the light-hole exciton state creates radiative excitons, which can either rapidly relax with the emission of photons or be ejected to the reservoir. The exciton scattering to the reservoir can be caused by the acoustic-phonon emission as well as by their interaction with each other and the excitons and carriers in the reservoir. The approximation of the dynamic curves for a weak excitation shown in Fig.~\ref{Dynamics1} gives a characteristic time of the ejection process of about 40~ps. This time is about twice longer than the radiative lifetime of the light-hole excitons. Consequently, only 1/3 fraction of the photocreated exciton popoulation is scattered to the reservoir. 

The interaction between the radiative excitons governs the time-dynamics of the nonradiative broadening on a short time scale shown in the inset of Fig.~\ref{Dynamics1}. We do not model this pulse as well as the ejection process discussed above. The excitons scattered to the reservoir can relax to the average reservoir temperature (fraction $k$ of the ``cold'' excitons) or dissociate into electrons and holes (the $(1-k)$ fraction of the ``hot'' excitons). We have to assume the exciton dissociation to explain the initial slow rise of the nonradiative broadening during several hundreds of picoseconds observed in the experiment at the strong excitation, see Fig.~\ref{Dynamics1}. The rate of this process is denoted as $\gamma_d$.
\begin{figure}
   \includegraphics{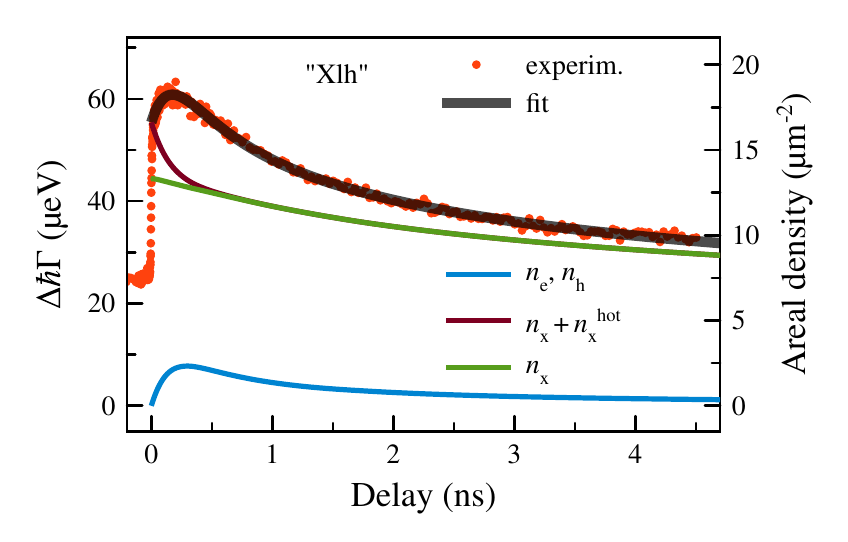}
   \caption{Dynamics of the Xhh nonradiative broadening under excitation to the ``Xlh'' resonance (points, left vertical axis) and its modeling (thick solid curve). The thin solid curves show the dynamics of the carrier and exciton densities (right vertical axis). Excitation power $P = 0.2$~mW.
   }
   \label{Fig6}
\end{figure}
The last two processes, which we take into account, are the electron-hole coupling into the exciton described by the rate $\kappa_{ex}$ (the bimolecular process~\cite{Piermarocchi-PRB1997}) and the carrier-induced exciton scattering from the reservoir to the light cone characterised by the bimolecular rate constant $\kappa_c$. For simplicity, we do not consider any phonon-induced processes assuming that they can be taken into account by the appropriate choice of the model parameters. We also do not consider any contribution of the exciton-exciton interaction to the scattering of the excitons from the reservoir to the light cone. The presence of the long-lived excitons in the reservoir indicates that this process should be less efficient than the exciton-carrier scattering.

The system of the kinetic equations describing the discussed processes are:
\begin{equation}
   \begin{aligned}
      n_x^{hot}       & = N_0(1-k)e^{-\gamma_d t},                        \\
      \frac{dn_e}{dt} & = \gamma_d n_x^{hot} - \kappa_{ex} n_e n_h,     \\
      \frac{dn_h}{dt} & = \gamma_d n_x^{hot} - \kappa_{ex} n_e n_h,     \\
      \frac{dn_x}{dt} & = \kappa_{ex} n_e n_h - \kappa_c n_x (n_e+n_h).
   \end{aligned}
   \label{Eqn-system1}
\end{equation}
Here $N_0$ is areal density of the excitons scattered to the reservoir; $n_e$, $n_h$, $n_x$, and $n_x^{hot}$ are, respectively, the electron, hole, and ``cold'' and ``hot'' exciton densities in the reservoir. The initial conditions for these variables are: $n_e(0) = n_h(0) = 0$,  $n_x(0) = N_{bgr} + N_0 k$, where $N_{bgr}$ is the background exciton density accumulated from the previous laser pulses.

This nonlinear system of equations does not allow for an analytical solution. We solved it numerically using the constants $N_0$, $k$, $\gamma_d$, $\kappa_{ex}$, and $\kappa_c$ as fitting parameters. The excitation-induced nonradiative broadening of the Xhh resonance is expressed via the carrier and exciton densities as follows:
\begin{equation}
   \Delta \hbar \Gamma_{\text{NR}} = \sigma_{eh} (n_e + n_h) + \sigma_x (n_{x}^{hot} + n_x).
   \label{scattering}
\end{equation}
Here $\sigma_{eh}$ and $\sigma_{x}$ are the cross-sections of the exciton-carrier (X-eh) and exciton-exciton (X-X) scatterings, respectively. The main mechanism of scattering of excitons and carriers at relatively small wave vectors is their exchange interaction, see Refs.~\cite{Ciuti-PRB1998, Cohen-PRB2003, Portnoi-EurPhysJ2008}. The analysis performed in these works shows that the exchange constants for X-eh and X-X interactions in narrow QWs have similar values. 
For simplicity, we put $\sigma_{eh} =\sigma_{x} \equiv \sigma$.

Examples of the theoretical curves obtained in this model are shown in Fig.~\ref{Fig6}. Thin solid lines in this panel show the dynamics of excitons and carriers in the reservoir at the excitation power $P_{exc} = 0.2$~mW. The dynamics equations for the electrons and holes are similar, see Eqs.~(\ref{Eqn-system1}), therefore $n_h(t) = n_e(t)$. 
The modeling shows that a fraction of ``cold'' excitons is of about $k \approx 0.7$ at this excitation power.
The obtained values of the bimolecular rate constants are $\kappa_{ex}= 0.6$~$\mu$m$^2/$ns and $\kappa_{c} = 0.08$~$\mu$m$^2/$ns. The values of $\kappa_{ex}$ are of the same order of magnitude as those reported in Refs.~\cite{Piermarocchi-PRB1996, Piermarocchi-PRB1997, Deveaud-ChemPhys2005}. The parameter $\kappa_{c}$ is of about one order of magnitude smaller than $\kappa_{ex}$, therefore the carrier-induced scattering of reservoir excitons to the light cone is the slowest process in our model. 

To calibrate the exciton and carrier densities created in the reservoir by a single laser pulse, we have estimated the number of the photocreated excitons for the case of weak excitation with a power $P_{exc} = 0.1$~mW. The number of photons that reach the QW layer from a single laser pulse at this power is: $N_{\text{phot}} \approx 3\times10^6$. 
The absorption coefficient for an exciton transition is calculated using a standard formula~\cite{Ivchenko-book2004}:
$$
   a(\omega) = \frac{2\Gamma_R\Gamma_{NR}}{(\omega-\omega_X)^2+(\Gamma_R + \Gamma_{NR})^2}.
$$
Taking into account the values of the radiative broadening of the Xlh resonance, $\hbar \Gamma_R \approx 15$~$\mu$eV, and its nonradiative broadening just before the pulse arrival, $\hbar\Gamma_{NR}\approx 100$~$\mu$eV, we obtain the maximum value of $a(\omega_X) \approx 0.23$. We should also take into account that only $20\%$ of the incident photons are absorbed because of the relatively large spectral width of the pump pulses (HWHM = 480~$\mu$eV) compared to that of the Xlh resonance ($\hbar(\Gamma_{\text{R}}+\Gamma_{\text{NR}}) = 115$~$\mu$eV) and also the presence of oscillating wings of the pulses taking 18\% of their energy. Accounting for the laser spot area on the sample, $S_{\text{spot}} \approx 7100$~$\mu$m$^2$, we obtain the areal density of the bright excitons created by a single laser pulse: $N_{\text{br}} = 20$~$\mu$m$^{-2}$. Approximately 1/3 of the photocreated excitons are ejected into the reservoir. 
Therefore, the initial exciton density scattered to reservoir is: $N_0 \approx 7$~$\mu$m$^{-2}$. The obtained value of the exciton density allows us to estimate the cross-section of the exciton-exciton scattering at this excitation power using expression~(\ref{scattering}): $\sigma_x \approx 3$~$\mu\text{eV}\cdot\mu\text{m}^2$.

\subsection{Nonresonant excitation}
The broadening dynamics of the Xhh resonance was experimentally measured at the excitation to different spectral points above the Xlh exciton transition, see Sect.~\ref{dynamics-experiment}. The excitation power was 0.2~mW.
%
%
\begin{figure}
   \includegraphics{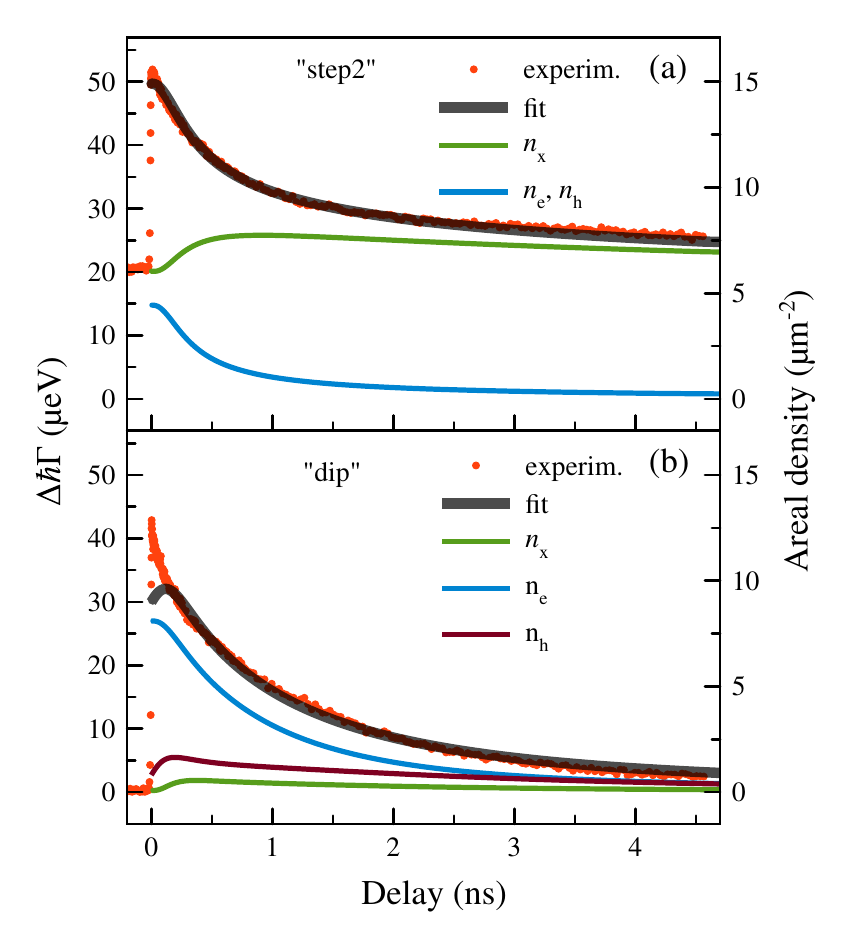}
   \caption{(a) Dynamics of the Xhh nonradiative broadening under excitation to the ``step2'' resonance (points, left vertical axis) and its modeling (thick solid curve). The thin solid curves show the dynamics of the carrier and exciton densities (right vertical axis). Excitation power $P = 0.2$~mW. (b) The same for the case of the excitation to the ``dip'' spectral region.
   }
   \label{Fig7}
\end{figure}
The excitation to the ``step1'', ``X3'', and ``step2'' spectral regions creates mainly free carriers in the QW. We put $n_x^{hot}=0$ in Eqs.~(\ref{Eqn-system1}) for simplicity. We also assume that the excitation creates hot electrons and holes, which cannot be coupled into excitons before their cooling. The cooling process is assumed to be exponential, $n_e^{hot} = n_h^{hot}  = N_0 e^{-\gamma_T t}$, with a characteristic time $\tau_T = 1/\gamma_T = 100$~ps. The equations describing the dynamics of the exciton and carrier densities are:
\begin{equation}
   \begin{aligned}
      \frac{dn_e}{dt} & = \gamma_T n_e^{hot} - \kappa_{ex} n_e n_h,     \\
      \frac{dn_h}{dt} & = \gamma_T n_h^{hot} - \kappa_{ex} n_e n_h,     \\
      \frac{dn_x}{dt} & = \kappa_{ex} n_e n_h - \kappa_c n_x (n_e+n_h).
   \end{aligned}
   \label{Eqn-system2}
\end{equation}
The initial condition for the exciton and carrier densities, $n_e(0)=n_h(0)=0$, $n_x(0) = N_{bgr}$. 

The fitting curve and the dynamics of the carrier and exciton densities obtained using Eqs.~(\ref{Eqn-system2}) are shown in Fig.~\ref{Fig7}(a). One can see that the theory well reproduces the general behavior of the dynamics of the nonradiative broadening. The carrier density considerably decreases during the first nanosecond due to the association of electrons and holes into excitons. The exciton density, after the initial increase, decays very slowly due to the single relaxation process: exciton scattering to the light cone caused by their interaction with free carriers. Having in mind that the carrier density is very small at $t > 1$~ns, the scattering process is slow. The obtained values of the bimolecular rate constants, $\kappa_{ex}= 0.9$~$\mu$m$^2/$ns and $\kappa_{c} = 0.07$~$\mu$m$^2/$ns. These values are close to those obtained at the weak excitation to the Xlh resonance.


The case of the excitation to the ``dip'' spectral region requires particular attention. In this case, one type of the carriers, presumably holes, can be created in the barrier layers and other one in the QW layer. As a result, an imbalance of the carriers of one sign appears in the QW. This imbalance can persist for a long time, while the carrier created in the barriers relax to the QW layer. This process was already illustrated in Fig.~\ref{Dynamics4}(b). Accordingly, the exciton reservoir can be efficiently depleted. The rate equations describing these processes slightly differ from Eqs.~(\ref{Eqn-system2}):
\begin{equation}
   \begin{aligned}
      \frac{dn_h}{dt} & = \gamma_b n_b + \gamma_T n_h^{hot} - \kappa_{ex} n_e n_h, \\
      \frac{dn_e}{dt} & =  \gamma_T n_e^{hot} - \kappa_{ex} n_e n_h,               \\
      \frac{dn_x}{dt} & = \kappa_{ex} n_e n_h - \kappa_c n_x (n_e+n_h).
   \end{aligned}
   \label{Eqn-system3}
\end{equation}
Here $n_b = N_0 (1-k)e^{-\gamma_b t}$, $n_e^{hot} =N_0 e^{-\gamma_T t}$, and $n_h^{hot}=k N_0 e^{-\gamma_T t}$. This system of equations is written for the case where the major part of holes [$(1-k)=0.9$ in this case] is created in the barriers. Consequently, the excess carriers in the QW are the electrons. The initial conditions for the exciton and carrier densities are: $n_h(0) = n_e(0) = 0$, $n_x(0) = N_{bgr}$.
The dynamics of the nonradiative broadening obtained in the modeling well reproduces the experimentally observed dependence, see Fig.~\ref{Fig7}(b). The only exception is the initial part of the dependence at $t < 150$~ps, which is not reproduced by the modeling. Possibly, some fast processes not considered in our model affect the broadening at the early stage of the carrier relaxation.
The areal density of electrons in the QW is relatively large at the initial time moment, $n_e(0) + n_e^{hot}(0)=8$~$\mu$m$^{-2}$, and slowly decays in time. The decay is mainly determined by the capture of holes from the barriers (characteristic time $t_b = 1/\gamma_b = 770$~ps) followed by their coupling with the electrons. The bimolecular rate constant for the coupling, $\kappa_{ex}= 0.77$~$\mu$m$^2/$ns, that is slightly larger than that for the case of the excitation to the Xlh resonance.

The scattering of the reservoir excitons to the light cone is very efficient in this case. This directly follows from the experimental data and their interpretation in the framework of our model.  The bimolecular rate constant for the exciton scattering is relatively large, $\kappa_{c} = 1.7$~$\mu$m$^2/$ns, at least one order of the magnitude larger than that obtained for the case of the excitation to the Xlh resonance. This can be partially explained by the low effective temperature of electrons, which are created in the QW and have enough time for cooling. There is possibly one more reason (or several reasons) for so efficient depletion of the exciton reservoir. We should mention that the experimentally observed dynamics of the exciton reservoir at the  stronger excitation (not shown here) becomes slower so that the nonzero broadening at the negative delay appears again. Such dynamics is characterized by a lower value of the rate $\kappa_{c}$.

\section{PL kinetics}
\label{PLkinetics}

The model described in the previous section predicts also the time dependence of the PL signal. Indeed, the PL profile is described by the last term in the last equation of systems~(\ref{Eqn-system1}, \ref{Eqn-system2}, \ref{Eqn-system3}),
\begin{equation}
   I_{\text{PL}}(t) = A_{\text{PL}} \kappa_c n_x (n_e + n_h),
   \label{PL-model}
\end{equation}
where $A_{\text{PL}}$ is a scaling factor.

\begin{figure*}
   \includegraphics{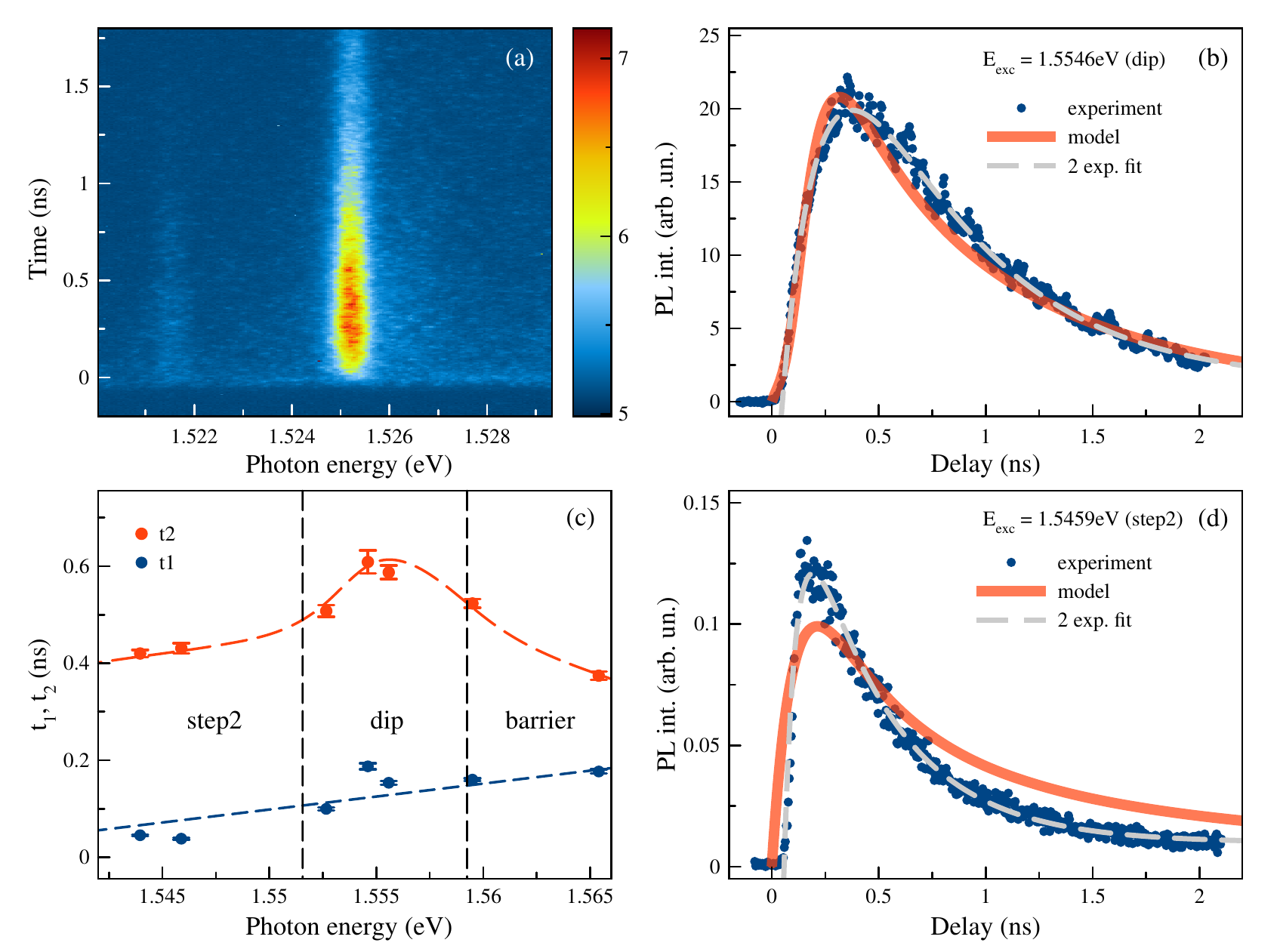}
   \caption{(a) An example of the color plot of the PL kinetics in the spectral region around of the Xhh exciton measured at the excitation to the ``dip'' region ($E_{exc} = 1.5546$~eV). Excitation power, $P_{exc} = 0.2$~mW; sample temperature, $T = 8.3$~K. Panels (b) and (d) show the cross-sections of the color plots measured at the excitation to the ``dip'' (the upper panel) and ``step2'' (the lower panel) regions. The dots are the experimental data. The dashed curves are the fits by function~(\ref{Eq-fit}). The solid curves are the modeling    by Eq.~(\ref{PL-model}). (c) The rise and decay times of the PL kinetics obtained at the excitation to different spectral points. The blue dashed line is a linear approximation. The red dashed curve is a guide to the eye.
   }
   \label{Fig-PL}
\end{figure*}

To verify the prediction by Eq.~(\ref{PL-model}), we have measured the kinetics of the Xhh exciton PL at different energies of photoexcitation. The kinetics was measured using a setup, which included a 2-ps tunable Ti:sapphire laser, a cryostat, a spectrometer, and a synchro-scan streak-camera. Examples of the PL kinetics are shown in Fig.~\ref{Fig-PL}. The left top panel shows a color map of the kinetics taken under excitation of the sample to the ``dip'', $E_{exc} = 1.5546$~eV. The two right panels show the cross-sections of similar color maps at the energy of Xhh exciton transition ($E_{Xhh} = 1.5252$~eV) measured at the excitation to the ``dip'' (the top panel) and to the ``step2'' (the bottom panel). The time dependencies of the PL can be approximated by a phenomenological two-exponential function~(\ref{Eq-fit}) with characteristic times $t_1$ and $t_2$, whose dependencies on the excitation photon energy are shown in the left bottom panel.

One can see that the PL signal is not vanishing at least during 2~ns after the excitation pulse. This is orders of magnitude longer than the radiative lifetime of the Xhh exciton, $\tau_{Xhh} = 1/(2\Gamma_R) \approx 10$~ps. Here we used the value of the radiative broadening, $\hbar\Gamma_{\text{R}} = 37$~$\mu$eV (see sect.~\ref{experimental}) and the well-known relation between the radiative time and $\Gamma_R$~\cite{Ivchenko-book2004}. Hence, the PL kinetics is governed by the relaxation processes in the nonradiative reservoir discussed in the previous sections.

The model of the dynamic processes in the reservoir reasonably well describes the PL pulse profile with no fitting parameters. This is a clear indication that all valuable processes are taken into account in the model. In particular, the PL signal decays almost down to zero at $t>2.5$~ns points out that the main mechanism of the scattering of the excitons from the reservoir to the light cone is governed by the exciton interaction with free carriers. The contribution of the exciton-exciton and exciton-phonon scattering is considerably smaller otherwise we would observe the PL ``tail'' all the time while the excitons are stored in the reservoir that is until the next laser pulse.

Finally, we should note that the PL data, similar to the dynamics of the Xhh broadening in the reflectivity spectra, show that the ``dip'' spectral region of the optical excitation differs from other regions. Indeed, as it is seen in Fig.~\ref{Fig-PL}(c), the PL decay time is characterized by the largest value of $t_2$ if the structure is excited to this spectral region. This is an intuitively understandable effect. The optical generation of excess carriers of one sign in the reservoir under excitation to the ``dip'' discussed in the previous section  supplies the PL channel by excitons all the time while the reservoir is not totally depleted. In other cases if the electron and hole densities are balanced, the ejection of reservoir excitons into the light cone slows down as soon as free electrons and holes disappear via association into excitons. At low temperatures, this happens long before the reservoir is depleted.

\section{Discussion}

The model of dynamic processes in the reservoir discussed in Sect.~\ref{Kinetics-model} allows us to quantitatively describe the dynamics of the nonradiative broadening of the Xhh resonance at different photon energies of excitation. This model also predicts a PL pulse profile, which should be observed in the experiment. The direct comparison of theoretical calculations with the experimental data shows that the model is able to correctly describe the PL pulse with no fitting parameters.

There are several points in the model which, ideally, should be verified. First, we assume one single universal constant $\sigma$, the cross-section of the exciton-carrier and exciton-exciton scatterings (see Eq.~(\ref{scattering}) and related text). This assumption applies, in particular, to the carriers in any energy states. At the same time, the carriers created in the barrier layer under excitation to the ``dip'' or ``barrier'' spectral regions, can acquire a large in-plane wave vector when they are captured in the QW layer. The exchange interaction of such carriers with excitons can be strongly changed as a theoretical modeling shows~\cite{Ciuti-PRB1998, Cohen-PRB2003, Portnoi-EurPhysJ2008}. Such carriers still can efficiently broaden the exciton resonances but they cannot play an important role in the scattering of the reservoir excitons into the light cone.

Second, the exciton-exciton scattering, which we ignored, can play some role in the depletion of the nonradiative reservoir. Our experimental data obtained for the resonant excitation to the Xlh transition with the low excitation density (see Fig.~\ref{Dynamics1}) when almost no free carriers are created clearly shows that the Xhh nonradiative broadening decays in time. This means that the exciton density in the reservoir decreases. Of course, this density decrease can be also caused by the exciton-phonon interaction~\cite{Piermarocchi-PRB1996, Piermarocchi-PRB1997}. It is desirable to include these processes in the model. However, to diminish the number of free parameters, the cross-section of these processes should be evaluated theoretically for the structure under study.

The presence of the reservoir of nonradiative excitons is able to explain the rapid increase of the PL intensity just after the excitation pulse observed in many works, see, e.g., Refs.~\cite{Szczytko-PRL2004, Deveaud-PRB2005, Deveaud-ChemPhys2005, Kaindl-PRB2009, Zybell-APL2014, Beck-PRB2016}. Indeed, the photocreated free carriers can scatter into the light cone the nonradiative excitons accumulated from previous pulses. An example of such a rapid increase of the PL is shown in Fig.~\ref{Fig-PL}(d). If the reservoir is empty, as in the case of excitation to the ``dip'', the PL rise is relatively slow, see Fig.~\ref{Fig-PL}(b).

Finally, we should mention that the rapid PL increase was a strong argument in favor of the model by Kira {\it et al.}~\cite{Kira-PRL1998}, in which this effect is explained by the resonant PL of the correlated electron-hole plasma. Our results show that the PL rise is naturally explained by the presence of real excitons in the nonradiative reservoir. A similar conclusion was achieved in Ref.~\cite{Kaindl-PRB2009}.

\section{Control of exciton reservoir}
The experimental results and the modeling discussed above show that, in high-quality QWs, the main mechanism of depleting the exciton reservoir is related to the exciton-free carrier scattering. This fact opens up an interesting possibility to control the exciton reservoir. We can populate the reservoir with excitons using optical excitation to one of the exciton resonances. On the other hand, we can populate the reservoir with uncompensated free carriers using another optical excitation to the ``dip’’ spectral region. If the areal density of free carriers is sufficient for the efficient depletion of the exciton reservoir, we should observe a {\it decrease} rather than an increase of the nonradiative broadening of exciton resonances when both the excitation beams are switched on.

\begin{figure}
   \includegraphics{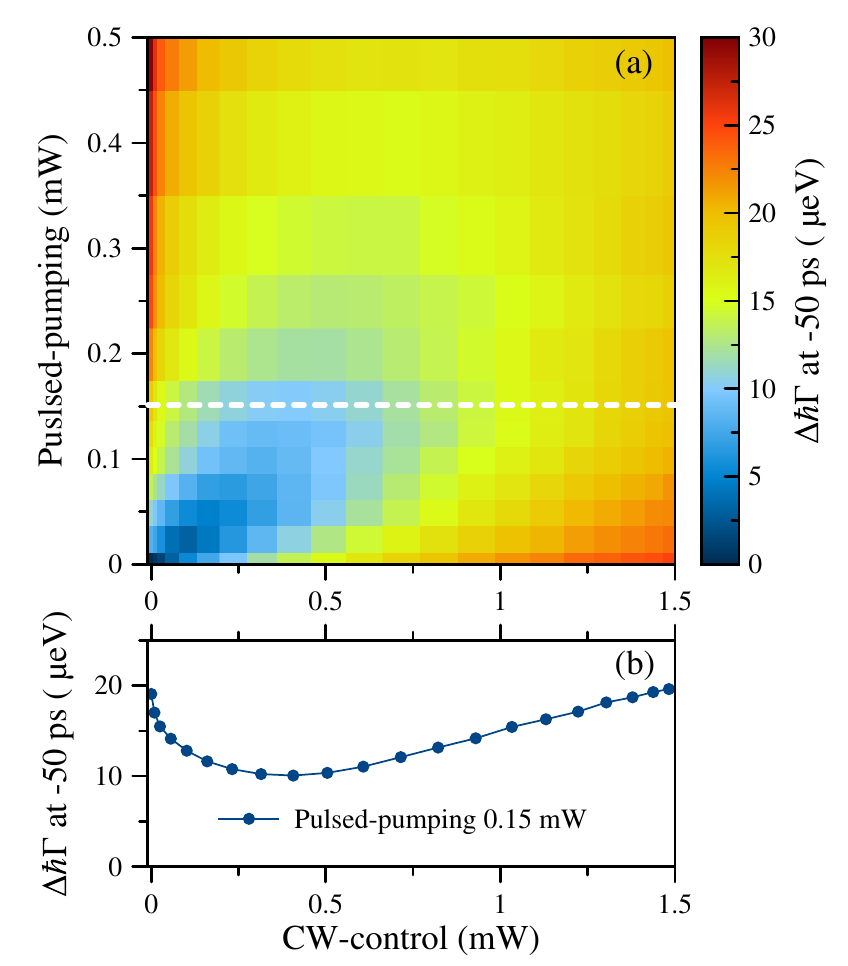}[h]
   \caption{(a) Color plot of the Xhh nonradiative broadening measured at small negative delays. The ``CW-control’’ is adjusted to the ``dip’’ spectral region and the ``Pulsed-pump’’ to the Xlh resonance. Panel (b) shows the cross-section of the panel (b) along the white dashed line.}
   \label{Fig9}
\end{figure}
\begin{figure}
   \includegraphics{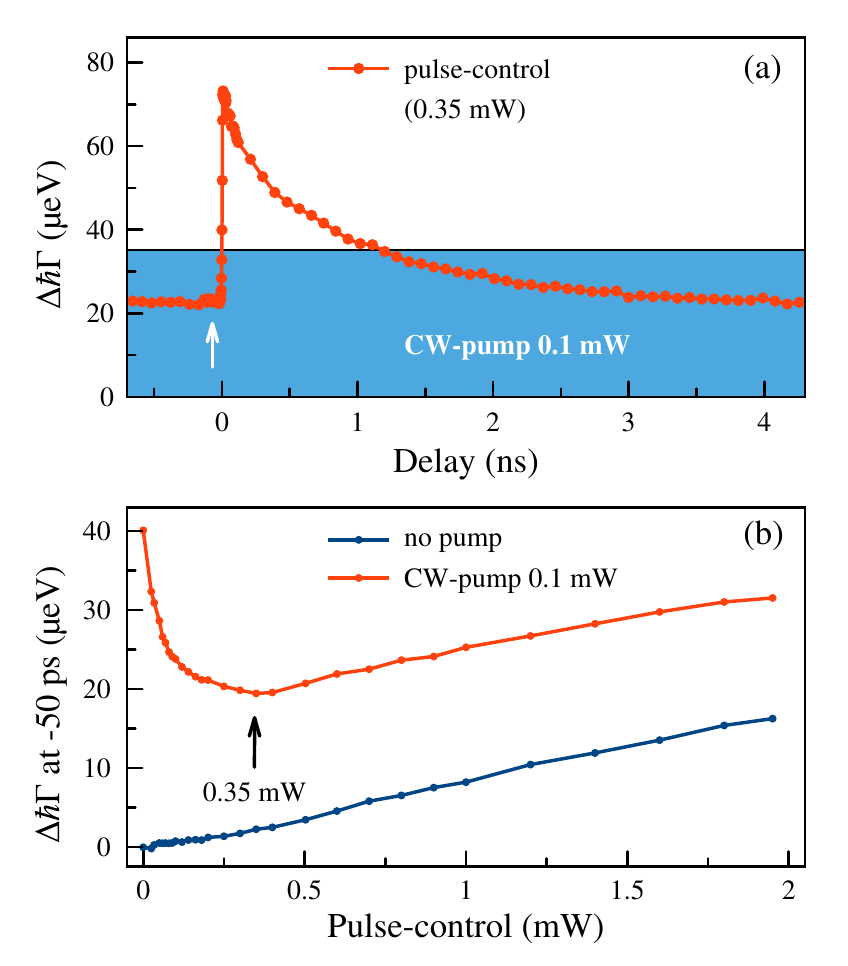}[h]
   \caption{(a). Dynamics of the Xhh nonradiative broadening under ``CW-pump’’ to the Xlh resonance ($P_{\text{CW}} = 0.1$~mW) and ``Pulsed-control’’ ($P_{pulse}=0.35$~mW) to the ``dip’’ spectral region. Blue area indicates the broadening without the ``Pulsed-control’’. (b) The ``Pulsed-control’’ power dependence of the Xhh nonradiative broadening measured at the negative delay in the presence of ``CW-pump’’ (red curve) and with no ``CW-pump’’ (blue curve).}
   \label{Fig10}
\end{figure}

We implemented this idea in two ways using the pulsed and CW laser beams. First, we use the laser pulses to create excitons in the reservoir via excitation to the Xlh resonance. We call this excitation ``pulsed-pumping''. We monitor the areal density of excitons in the reservoir, measuring the nonradiative broadening of the Xhh exciton resonance at the negative delay between the pump and probe pulses. The CW laser beam called ``CW-control’’ is used to create free carriers in the reservoir via excitation to the ``dip’’ spectral region. The two-dimensional plot shown in Fig.~\ref{Fig9}(a) demonstrates the results of the measurements when the ``CW-control’’ and ``Pulsed-pumping’’ excitation powers are scanned in some ranges. As seen, the ``CW-control’’ really allows one to deplete the exciton reservoir partially. An example of a cross-section of the two-dimensional plot shown in panel (b) of this figure demonstrates an approximately twofold decrease of the nonradiative broadening and, correspondingly, of the exciton areal density in the reservoir at the optimal ``CW-control’’ power of about 0.4 mW.

Second, we use the opposite experimental configuration when the ``CW-pump’’ creates the exciton reservoir via excitation to the Xlh resonance and the ``Pulsed-control’’ creates the unbalanced free carriers via excitation to the ``dip’’ spectral region. As seen in Fig.~\ref{Fig10}(a), the ``CW-pump’’ alone with power of 0.1~mW creates the exciton reservoir, which additionally broadens the Xhh resonance by about of 40~$\mu$eV (blue area in this figure). When the ``Pulsed-control’’ is switched on, the dynamics of the nonradiative broadening demonstrate an intense peak during the first one nanosecond (see the red curve in Fig.~\ref{Fig10}(a)). As discussed above (Sect~\ref{Kinetics-model}), this peak reveals due to the creation of the free carriers. At larger delays, the broadening becomes {\it smaller} than that in the absence of the ``Pulsed-control’’. This is a clear indication of depleting the exciton reservoir by free carriers created by the ``Pulsed-control’’. Panel (b) in this figure illustrates the power dependence of the nonradiative broadening at the negative delay. When no ``CW-pump’’ is present, the broadening and, correspondingly, the exciton density in the reservoir is relatively small (see blue curve). The red curve clearly shows the effect of depopulation of the exciton reservoir created by the ``CW-pump’’. Similar to the case shown in Fig.~\ref{Fig9}(b), the approximately twofold decrease of the nonradiative broadening is observed when the ``Pulsed-control’’ is switched on with some optimal power of about 0.35~mW.

\section{conclusion}

In this work, we experimentally studied the dynamics of nonradiative excitons with large in-plane $K$ vectors in high-quality structures containing a shallow QW. The main method we used is based on the detection of the nonradiative broadening of exciton resonances under the action of the CW or pulsed excitation. The CW excitation with different photon energies allows one to measure NBE spectra, which bring a valuable information about the energy spectra of excitons and free carriers. We found several spectral regions characterized by the different behaviors of the broadening of the Xhh and Xlh resonances. In particular, in the case of excitation into exciton resonances, as well as in the regions ``step1" and ``step2", the light-induced broadening of the Xlh resonance is approximately half of that of the Xhh resonance. We explain this phenomenon by the accumulation of the Xhh excitons, rather than Xlh ones, in the nonradiative reservoir. The radiative and nonradiative Xhh excitons interact via exchange by both electrons and heavy holes while the radiative Xlh--nonradiative Xhh exciton interaction occurs via the electron exchange only.

The CW excitation with higher photon energy reveals a ``dip'' spectral region followed by optical transitions in barrier layers. In  this regime, the nonradiative broadenings of the Xhh and Xlh resonances are equal and considerably smaller than in the other regions. The PLE spectra measured in similar experimental conditions show, however, no dip in the intensity under excitation to this spectral region. At the same time, the broadening of the Xhh PL line again demonstrates the ``dip'' region. These experimental observations allowed us to ascribe the ``dip'' region to the optical transitions between the states of free carriers localized in the QW and those delocalized in the barriers layer. Such optical transitions create an imbalance of carriers of one sign in the QW. This imbalance strongly affects the dynamics of nonradiative excitons in the reservoir. In particular, it gives rise to its efficient depletion.

Using a non-degenerated spectrally resolved pump-probe technique we studied the dynamics of the nonradiative excitons under the pulsed excitation to selected spectral regions. If the excitation created the excitons and free carriers in the QW layer, the dynamics is characterized by a fast decay of the nonradiative broadening of the Xhh and Xlh resonances followed by a long-lived ``tail'' surviving up to the next laser pulse. This is a clear indication of the accumulation of the nonradiative excitons in the reservoir. In the case of excitation to the ``dip'' spectral region, no ``tail'' is observed. The `` tail'' appears again if exciting with the higher photon energy corresponding to the optical transitions in the barrier layers.

We proposed a kinetic model describing the dynamics of the nonradiative excitons. To minimize the number of free parameters, we considered very few dynamic processes in the reservoir. Namely, the association of electrons and holes into excitons and the ejection of the excitons to the light cone are considered in the cases where the optical excitation created the free carriers in the QW layer. At the higher photon energy of the excitation, a capture of the photocreated carriers from the barrier layers to the QW layer is additionally considered. The model well describes the exciton dynamics observed experimentally. The obtained values of the bimolecular coupling rate are close to those reported in the literature.

To verify the model, we studied the PL kinetics in this structure. We found that the profile of the PL pulse is well described by the model without fitting parameters.

We also have demonstrated a possibility to manipulate the exciton reservoir areal density by the use of two optical beams exciting the sample into different spectral regions. Potentially, this effect can be used for creation of a desirable potential landscape for polaritons in the microcavity heterostructures. Such landscape is important for the deterministic control of coupling of the microcavity exciton-polariton condensates~\cite{Lagoudakis-PRL2020}.

\section*{ACKNOWLEDGMENTS}
The authors thank D. Mursalimov for technical assistance in the experiments. Financial support from SPbU, grant No. 73031758, and the Russian Foundation for Basic Research, grants No. 19-52-12032, No. 19-02-00576a, and No. 20-32-70131, is acknowledged. The authors thank Recourse Center ``Nanophotonics'' SPbU for the heterostructure studied in the present work. The authors thank the developers of the MagicPlot software, which was used to analyze a large amount of experimental data. Alexey Kavokin acknowledges the support by the Westlake University (Project No. 041020100118).

\end{document}